\documentclass[a4paper,11pt]{article}

\usepackage{pos}
\usepackage{amsmath}
\usepackage{slashed}
\usepackage[utf8]{inputenc}
\usepackage{graphicx}
\usepackage{subcaption}
\usepackage{hyperref}

\graphicspath{{./figures/}}

\newcommand{\Nf}{N_{\mathrm{f}}}
\newcommand{\Ns}{N_{\mathrm{s}}}
\newcommand{\Nt}{N_{\mathrm{t}}}
\newcommand{\ii}{\mathrm{i}}
\newcommand{\g}{\gamma_{*}}
\newcommand{\Cs}{C_{\mathrm{short}}}
\newcommand{\Cl}{C_{\mathrm{long}}}
\renewcommand{\L}{\mathcal{L}}

\title{Remnants of large-$\Nf$ inhomogeneities in the 2-flavor chiral Gross-Neveu model}
\ShortTitle{Remnant inhomogeneities in the 2-flavor chiral GN model}

\author{Julian J. Lenz}
\author*{Michael Mandl}

\affiliation{Theoretisch-Physikalisches Institut, Friedrich-Schiller-Universität Jena,\\
  Fröbelstieg 1, D-07743 Jena, Germany}

\emailAdd{julian.johannes.lenz@uni-jena.de}
\emailAdd{michael.mandl@uni-jena.de}

\abstract{We study the $(1+1)$-dimensional chiral Gross-Neveu model on the lattice. At finite density, analytic 
mean-field results predict the existence of inhomogeneous condensates breaking both chiral 
symmetry and spacetime symmetries spontaneously. We investigate the fate of these 
inhomogeneities for two flavors and find remnant structural order, albeit with a decaying 
amplitude. We also map out phase diagrams in the plane spanned by the chemical potential and 
temperature for different lattice spacings and physical volumes. Finally, we comment on the 
interpretation of our results in the light of various no-go theorems.}

\FullConference{%
 The 38th International Symposium on Lattice Field Theory, LATTICE2021
  26th-30th July, 2021
  Zoom/Gather@Massachusetts Institute of Technology
}

\begin{document}
\maketitle

\section{Introduction}
Despite tremendous efforts, only very little is known about the
phase diagram of Quantum Chromodynamics (QCD), the theory governing
the strong interactions, at high baryon density and low temperature.
This is because in this region first-principles lattice methods
cannot be applied due to the complex-action problem.

An alternative approach is to study effective theories 
or toy models, e.g.\@ of the Nambu-Jona-Lasinio (NJL) \cite{NJL61} or 
Gross-Neveu (GN) \cite{GN74} type. They are purely fermionic 
theories and, despite their apparent simplicity, share various 
important features with QCD, such as renormalizability (in low dimensions), asymptotic 
freedom, chiral symmetry and its potential spontaneous breakdown. It is the 
latter feature that we shall be concerned with in this 
contribution.

It has become clear \cite{ST00,ST01r,TU03} (see also \cite{Thi06} for
a review) that the (1+1)-dimensional versions of a number of GN- or 
NJL-type models exhibit inhomogeneous condensates, indicating the 
breakdown of a combination of chiral symmetry and spacetime symmetries,
at least for an infinite number of 
fermion flavors $\Nf$ (or on a mean-field level). Very recently it was
revealed by a large-scale 
lattice study that in the conventional $\mathbb{Z}_2$-symmetric GN 
model in $1+1$ dimensions a similar observation also holds for finite 
$\Nf$ \cite{LPW20,LPW20_1}. In this work we present an analogous 
study of the $U(1)$-symmetric chiral GN (cGN) model (see also \cite{LMW21}).

\section{The chiral Gross-Neveu model}
The cGN model is defined by its Lagrangian,
\begin{align}\label{eq:lagrangian_4F}
	\L = \bar{\psi}\ii\slashed{\partial}\psi +\frac{g^2}{2\Nf}\left((\bar{\psi}\psi)^2+(\bar{\psi}\ii\g\psi)^2\right)\;,
\end{align}
where $\psi$ implicitly combines $\Nf$ flavors of two-component spinors and $\g=i\gamma_0\gamma_1$ is 
the two-dimensional analogon of $\gamma_5$. In addition to spacetime and flavor 
symmetries the theory is invariant under continuous axial $U(1)$ 
transformations, 
\begin{equation}\label{eq:chiral_symmetry}
	\psi \rightarrow e^{\ii\alpha\g}\psi\;, \quad \bar{\psi} \rightarrow \bar{\psi}\,e^{\ii\alpha\g}\;, \quad \alpha\in\mathbb{R}\;.
\end{equation}

It is a common practice to trade the four-fermion terms in 
\eqref{eq:lagrangian_4F} for auxiliary scalar and pseudoscalar fields 
$\sigma$ and $\pi$, giving the equivalent Lagrangian
\begin{align}\label{eq:lagrangian}
	\L = \ii\bar{\psi}\left(\slashed{\partial}+\sigma+i\g\pi+\mu\gamma_0\right)\psi+\frac{\Nf}{2g^2}\left(\sigma^2+\pi^2\right)\;,
\end{align}
where we have also introduced a chemical potential $\mu$ in the 
usual way. The expectation values of these auxiliary fields are 
related via Dyson-Schwinger equations to the chiral and pseudoscalar 
condensates,
\begin{equation}\label{eq:dyson_schwinger}
	\langle\bar{\psi}\psi\rangle = \frac{\ii\Nf}{g^2}\langle\sigma\rangle\;,\quad  \langle\bar{\psi}\g\psi\rangle = \frac{\Nf}{g^2}\langle\pi\rangle\;.
\end{equation}

\begin{figure}[t]
	\centering
	\includegraphics[scale=01]{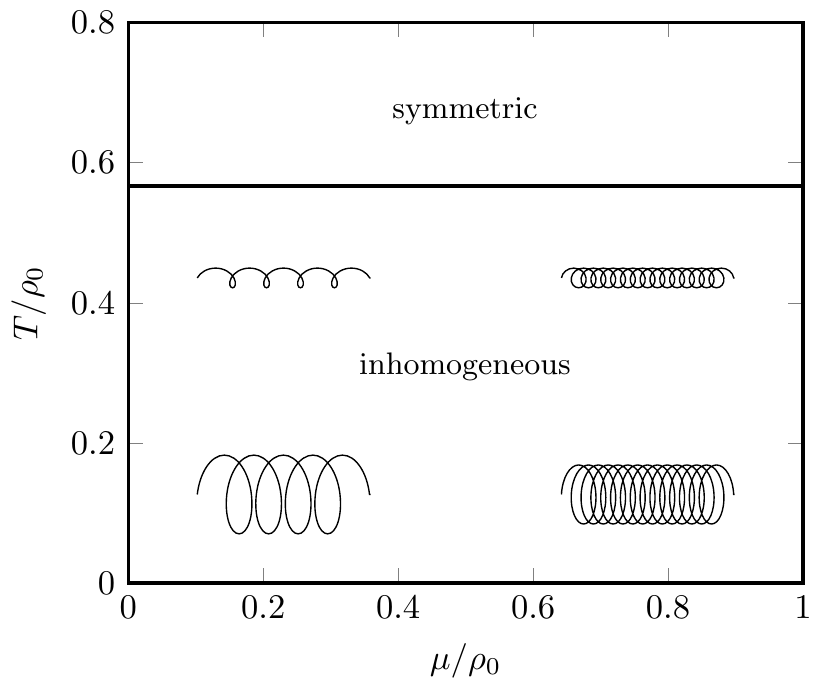}
	\caption{Large-$\Nf$ phase diagram showing a symmetric phase above 
	and an inhomogeneous chiral spiral phase below $T_c$, see also \cite{ST01r}.}
	\label{fig:pd_largeN}
\end{figure}

The cGN model has been solved analytically for finite $\mu$ and 
temperature $T$ in the limit $\Nf\rightarrow\infty$
\cite{ST00,ST01r}. It shows a $\mu$-independent critical
temperature $T_c\approx0.567\rho_0$, where 
$\rho_0=\langle\rho\rangle_{T=0=\mu}$ and $\rho^2=\sigma^2+\pi^2$ 
\citep{BDT09}. Above $T_c$ the theory is in a chirally symmetric 
phase with vanishing $\langle\bar{\psi}\psi\rangle$ and 
$\langle\bar{\psi}\g\psi\rangle$ , while below $T_c$ one finds 
inhomogeneous condensates,
\begin{equation}\label{eq:largeN_chiral_spiral}
	\langle\bar{\psi}\psi\rangle(x) = \rho(T)\cos(2\mu x)\;,\quad  \langle\bar{\psi}\g\psi\rangle(x) = -\rho(T)\sin(2\mu x)\;,
\end{equation}
which together are called the \emph{chiral spiral} \cite{ST00}. 
Its amplitude $\rho(T)$ depends on the temperature but not on 
the chemical potential while its wave number $k=2\mu$ is only 
$\mu$-dependent. We depict the large-$\Nf$ phase diagram in the 
$(\mu,T)$ plane in Fig.~\ref{fig:pd_largeN}.

This work, along with \cite{LMW21}, serves to answer the question 
whether the inhomogeneous condensates are purely a large-$\Nf$ 
artifact or if instead there is a remnant structural order 
even at finite flavor number, where quantum fluctuations 
are no longer suppressed. For the $\mathbb{Z}_2$-symmetric GN 
model this question has been answered in \cite{LPW20,LPW20_1}, 
where the inhomogeneities were shown to survive the continuum 
limit even for $\Nf=2$. There are, however, numerous no-go theorems 
\cite{MW66,Hoh67,Col73} that prohibit the spontaneous breaking 
of continous symmetries in low dimensions.
While the participation of external symmetries at finite density
invites subtle discussions about their applicability,
there is no doubt about their validity at vanishing chemical potential.
All in all, the interpretation of these results a rather delicate 
issue, c.f.\@ \cite{LPW20}.

\section{Correlators and related quantities}
We study the cGN model \eqref{eq:lagrangian} on a finite 
two-dimensional spacetime lattice with $\Nt$ and $\Ns$ points
in the temporal and spatial directions respectively, using the
chiral SLAC derivative \cite{DWY76_1,BKU08} as the fermion
discretization of choice. Since the previous works 
\cite{LPW20,LPW20_1} focused on the case $\Nf=8$, in which the 
results resembled the large-$\Nf$ findings in many ways, we shall 
be concerned with $\Nf=2$ in this contribution with the goal of better 
understanding the effect of fluctuations. In order to facilitate 
comparisons with large-$\Nf$ results we use the quantity
\begin{align}
	\rho_0=\left\langle\sqrt{\sigma^2+\pi^2}\right\rangle_{T\approx0=\mu}
\end{align}
to set the scale. We simulate lattices with different lattice spacings 
$a$ as well as spatial extents $L=\Ns a$ to study finite-size 
and discretization effects. We refer to previous works for details on
the lattice setup \cite{LPW20,LMW21}, implementation \cite{LWW19} and the exact scale-setting 
procedure \cite{LMW21}, the latter of which is quite a subtle affair.

\begin{figure}[t]
	\centering
	\includegraphics[scale=0.6]{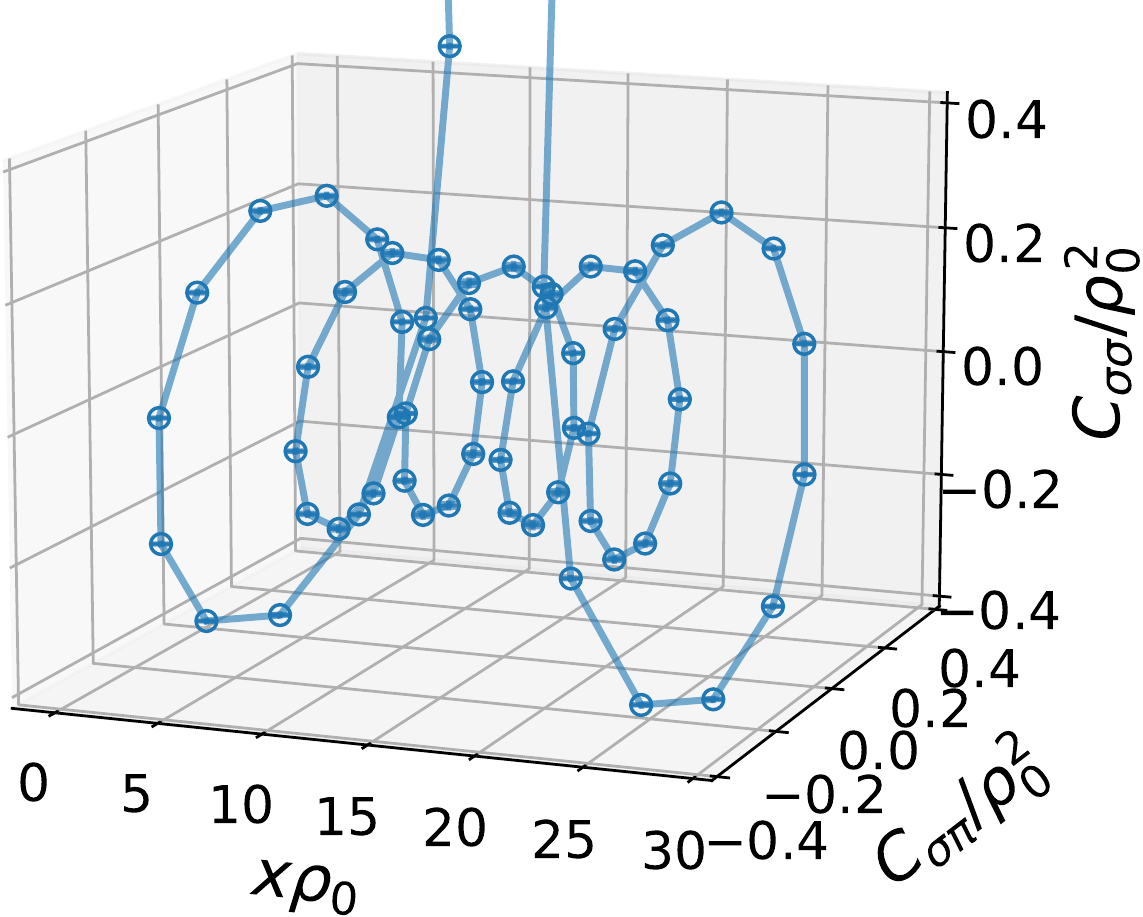}
	\caption{Chiral spiral on a lattice with $\Ns=63$, $T/\rho_0\approx0.030$,
	$\mu/\rho_0\approx0.88$ and $a\rho_0\approx0.46$.}
	\label{fig:chiral_spiral}
\end{figure}

It has proven useful in the past to probe for inhomogeneous phases by 
studying the spatial correlation functions of the auxiliary fields, 
\begin{align}\label{eq:spatial_correlators}
	\begin{aligned}
		C_{\sigma\sigma}(x)&=\frac{1}{N_tN_s}\sum_{t,y}\langle\sigma(t,y+x)\sigma(t,y)\rangle\;,\\
		C_{\sigma\pi}(x)&=\frac{1}{N_tN_s}\sum_{t,y}\langle\sigma(t,y+x)\pi(t,y)\rangle\;,
	\end{aligned}
\end{align}
where the sums run over all lattice points. 
Their use avoids cancellations due to the exact preservation of all symmetries
in the finite volume and instead pronounces the common structures of the generated 
Monte-Carlo configurations.

\begin{figure}[t]
\centering
\begin{subfigure}{.45\linewidth}
	\includegraphics[scale=0.4]{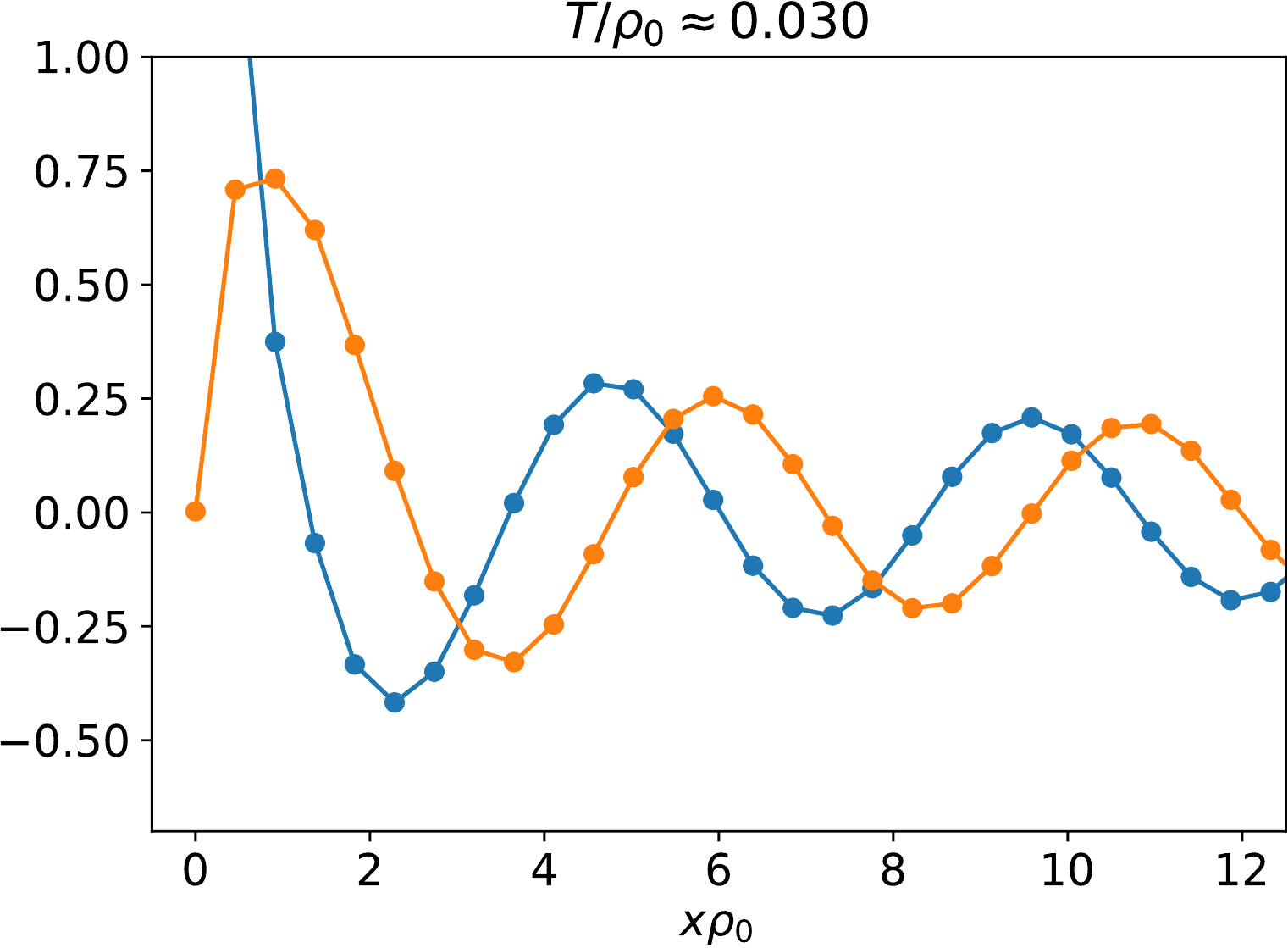}
\end{subfigure}
\begin{subfigure}{.45\linewidth}
	\includegraphics[scale=0.4]{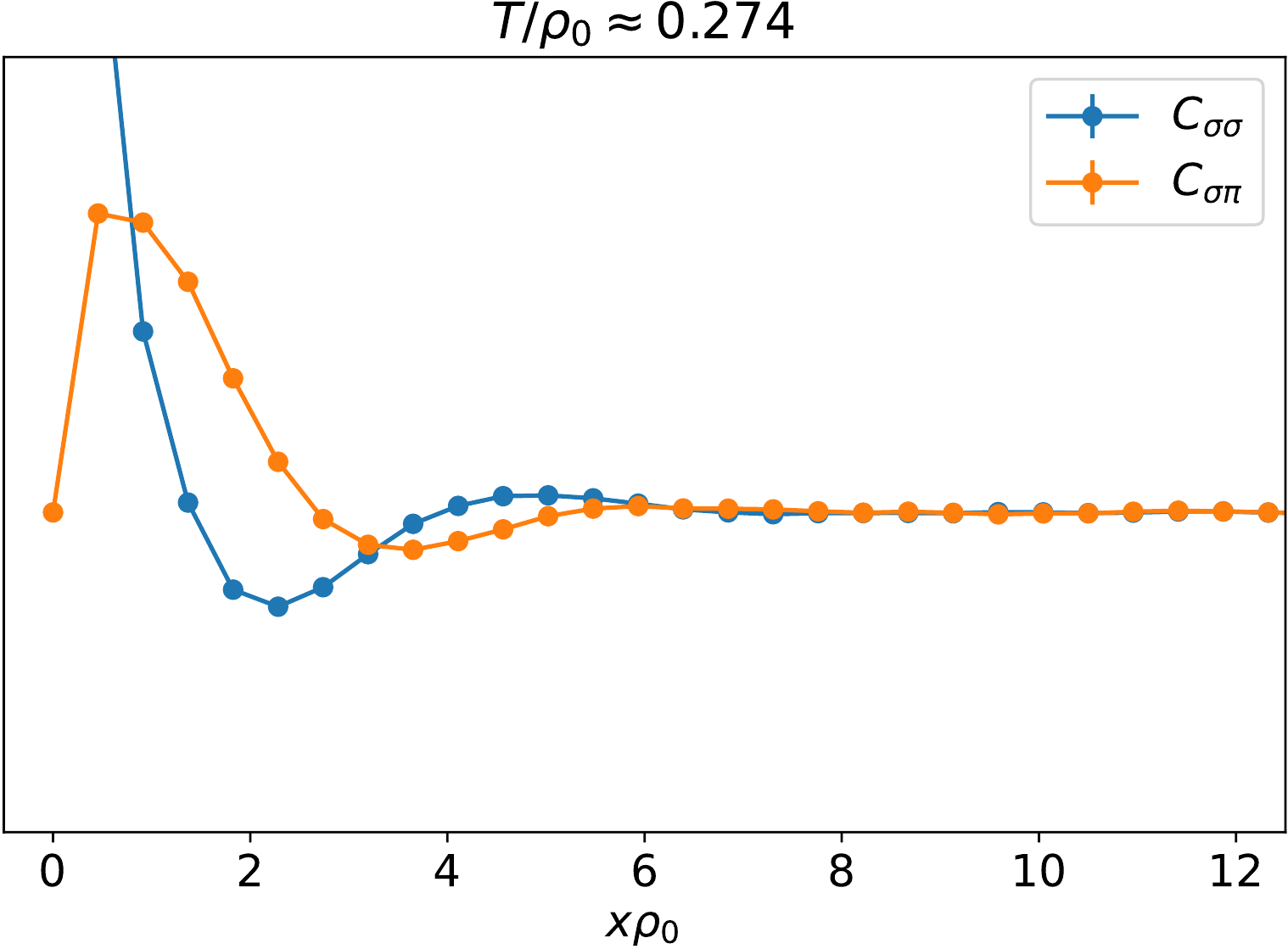}
\end{subfigure}
\caption{Spatial correlation functions $C_{\sigma\sigma}(x)$ and
	$C_{\sigma\pi}(x)$ on a lattice with $\Ns=63$, $\mu/\rho_0\approx0.88$ 
	and $a\rho_0\approx0.46$ for two different temperatures.}
\label{fig:correlators}
\end{figure}

We show a 3D plot of typical spatial correlators we find at low 
$T=1/\Nt a$ and intermediate to high $\mu$ in Fig.~\ref{fig:chiral_spiral}. 
We see that there is indeed structural order at finite $\Nf$. It is 
important to stress at this point that this is likely no perfect long-range 
order. Otherwise, that would imply spontaneous symmetry breaking which is 
prohibited by the no-go theorems mentioned above. Indeed, the amplitude of 
the chiral spiral decays with $x$ (see also the left plot of 
Fig.~\ref{fig:correlators}), indicating some sort of \emph{quasi-long-range} 
order. 

Analytical studies at vanishing \cite{And05} and finite \cite{LMW21} density 
find a massive phase at finite temperature that becomes critical at $T=0$.
Thus, we expect the correlators to decay
exponentially with a thermal mass that vanishes as $T\to 0$. 
To investigate this issue we show in Fig.~\ref{fig:correlators} 
a comparision between the spatial correlation functions at low and high 
temperature. Indeed, while still oscillatory, the correlators decay more 
rapidly as one increases the temperature supporting the analytical claims. 
Our results are also consistent with the "quantum spin liquid" scenario 
discussed in \cite{PTV20}. We remark that this decrease of correlation 
functions does not occur in the large-$\Nf$ limit and is thus caused by 
fluctuations (see also \cite{Wit78}).

Our main interest in this contribution is to study the phase 
diagram of the $(1+1)$-dimensional cGN model at $\Nf=2$. 
To condense an extensive study of the correlators \eqref{eq:spatial_correlators}
we propose to study the two quantities
\begin{align}\label{eq:cshort}
	\Cs &= \min_x C_{\sigma\sigma}(x)\;,
	\\
	\label{eq:clong}
	\Cl &= \min_x \sqrt{C_{\sigma\sigma}^2(x)+C_{\sigma\pi}^2(x)}\;.
\end{align}
In the limit of infinite flavor number 
non-vanishing values of these quantities would indicate spontaneous symmetry breaking.
In this scenario, the sign of $\Cs$ (which was denoted as $C_{\min}$ in \cite{LPW20,LPW20_1,LMW21})
would discriminate inhomogeneous from homogeneous spontaneous symmetry breaking.
For $\Nf=2$, where long-range order is most unlikely, one has to be more careful
in their interpretation: Whenever $\Cs$ or $\Cl$ are positive (beyond noise), the
system exhibits correlated patches of space that are of the order of the lattice volume.
In that sense, correlations are at least quasi-long-ranged if the signal persists in
the infinite-volume limit. A negative value of $\Cs$ indicates further that there are 
inhomogeneities on some scale. This scale, however, by no means has to be large,
as is nicely illustrated by the right-hand plot of Fig.~\ref{fig:correlators}.
In fact, negative values of $\Cs$ will always stem from the short-range behavior,
no matter if correlations persist on larger scales or not, because the correlations
are largest on short scales. We summarize this discussion in Tab.~\ref{t:interpretation_order}.

\begin{table}
	\renewcommand{\arraystretch}{1.5}
	\caption{\label{t:interpretation_order}Interpretation of the quantities \eqref{eq:cshort} and \eqref{eq:clong} for $\Nf=\infty$, with the possibility of spontaneous symmetry breaking, and for $\Nf=2$, where most likely no strict long-range order can exist.}
	\begin{tabular}{ccccc}
		\hline\hline
		&\multicolumn{2}{c}{$\Cs$}&\multicolumn{2}{c}{$\Cl$}
		\\
		\hline
		&$\Nf=\infty$&$\Nf=2$&
		$\Nf=\infty$&$\Nf=2$
		\\
		\hline
		$>0$&hom. broken&hom.\@ (quasi-)long-ranged&broken (any kind)&(quasi-)long-ranged\\
		$\approx 0$&symmetric&short-ranged&symmetric&short-ranged\\
		$<0$&inhom. broken& any inhomogeneities & -- & --
		\\
		\hline
		\hline
	\end{tabular}
\end{table}

One should note that neither $\Cs$ nor $\Cl$ is a local observable and it is therefore
questionable if they could be considered as legitimate order parameters. However, at finite 
$\Nf$ there is most likely no long-range order in the strict sense to be found.

\section{Phase diagrams}

\begin{figure}
\begin{subfigure}{.333\linewidth}
	\includegraphics[scale=0.33]{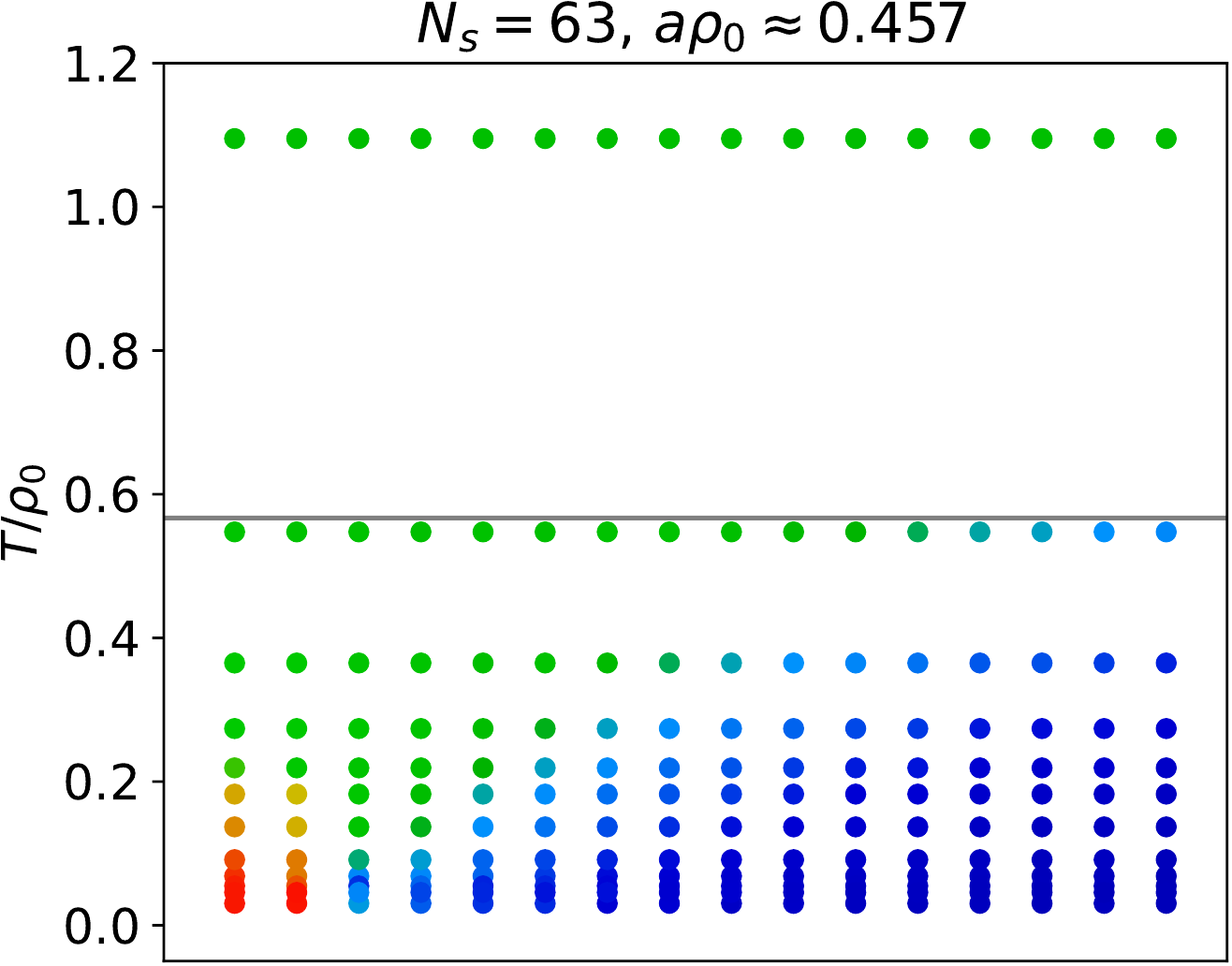}
\end{subfigure}
\hspace{-2.3cm}
\begin{subfigure}{.333\linewidth}
	\includegraphics[scale=0.33]{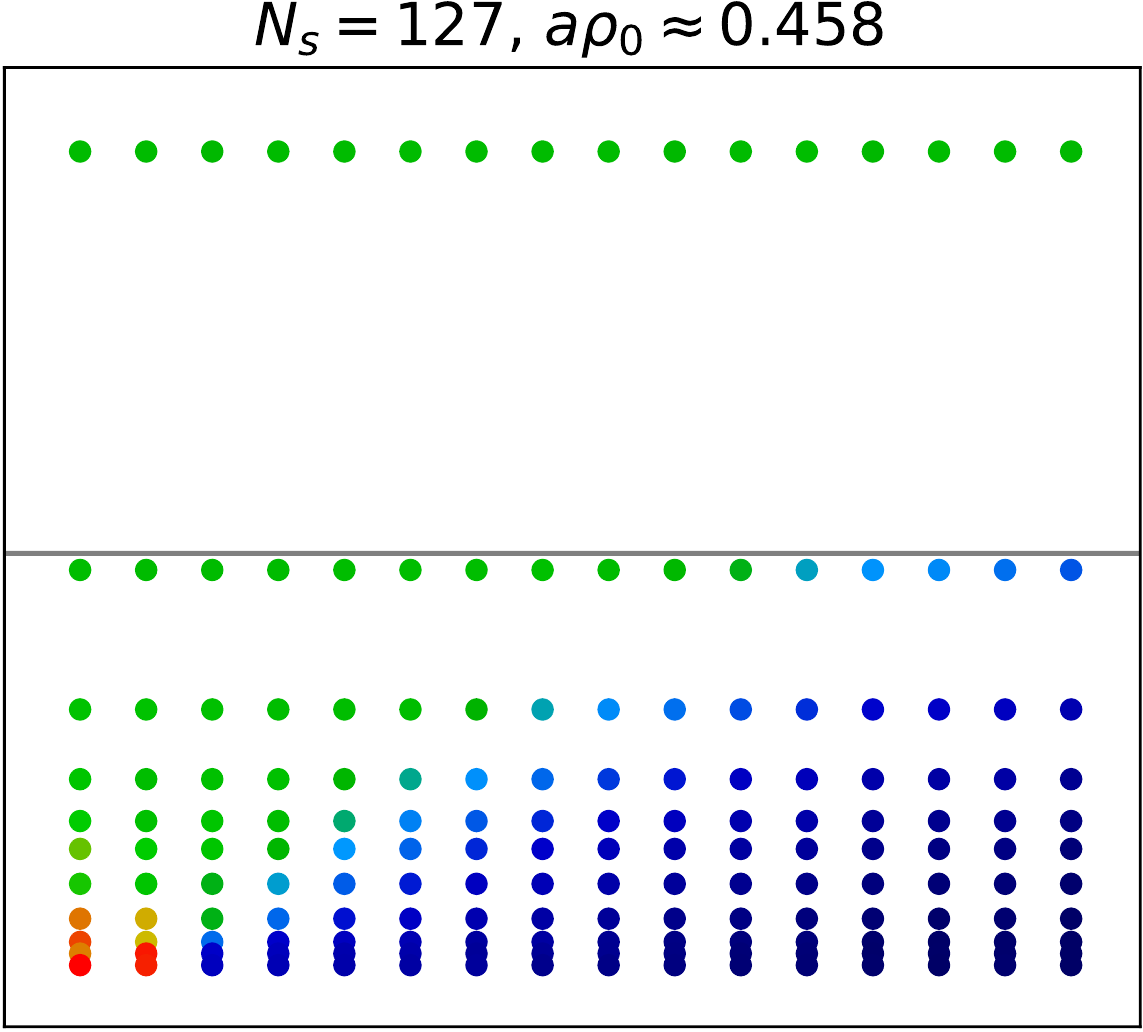}
\end{subfigure}
\hspace{-2.8cm}
\begin{subfigure}{.333\linewidth}
	\includegraphics[scale=0.33]{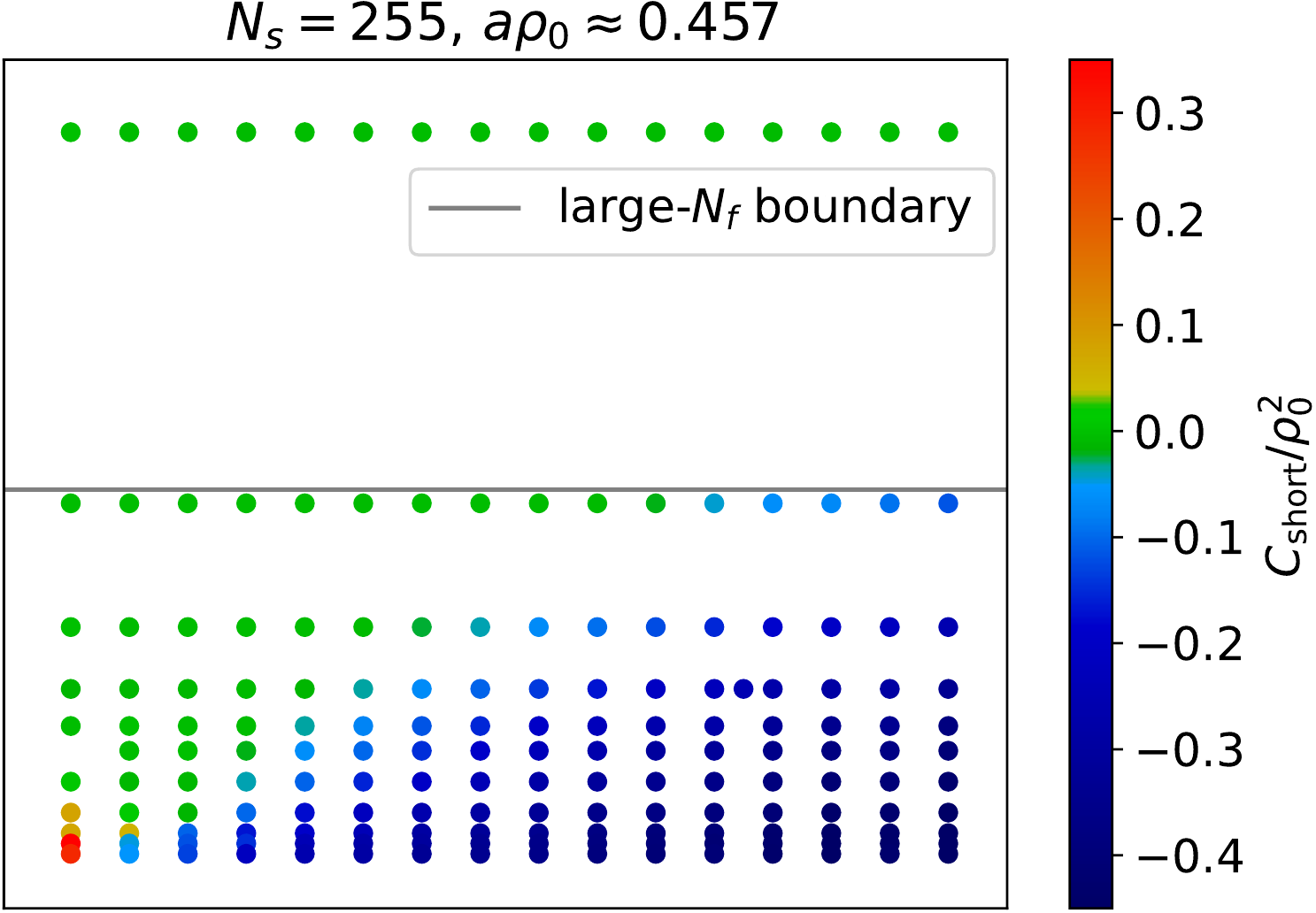}
\end{subfigure}
\begin{subfigure}{.333\linewidth}
	\includegraphics[scale=0.33]{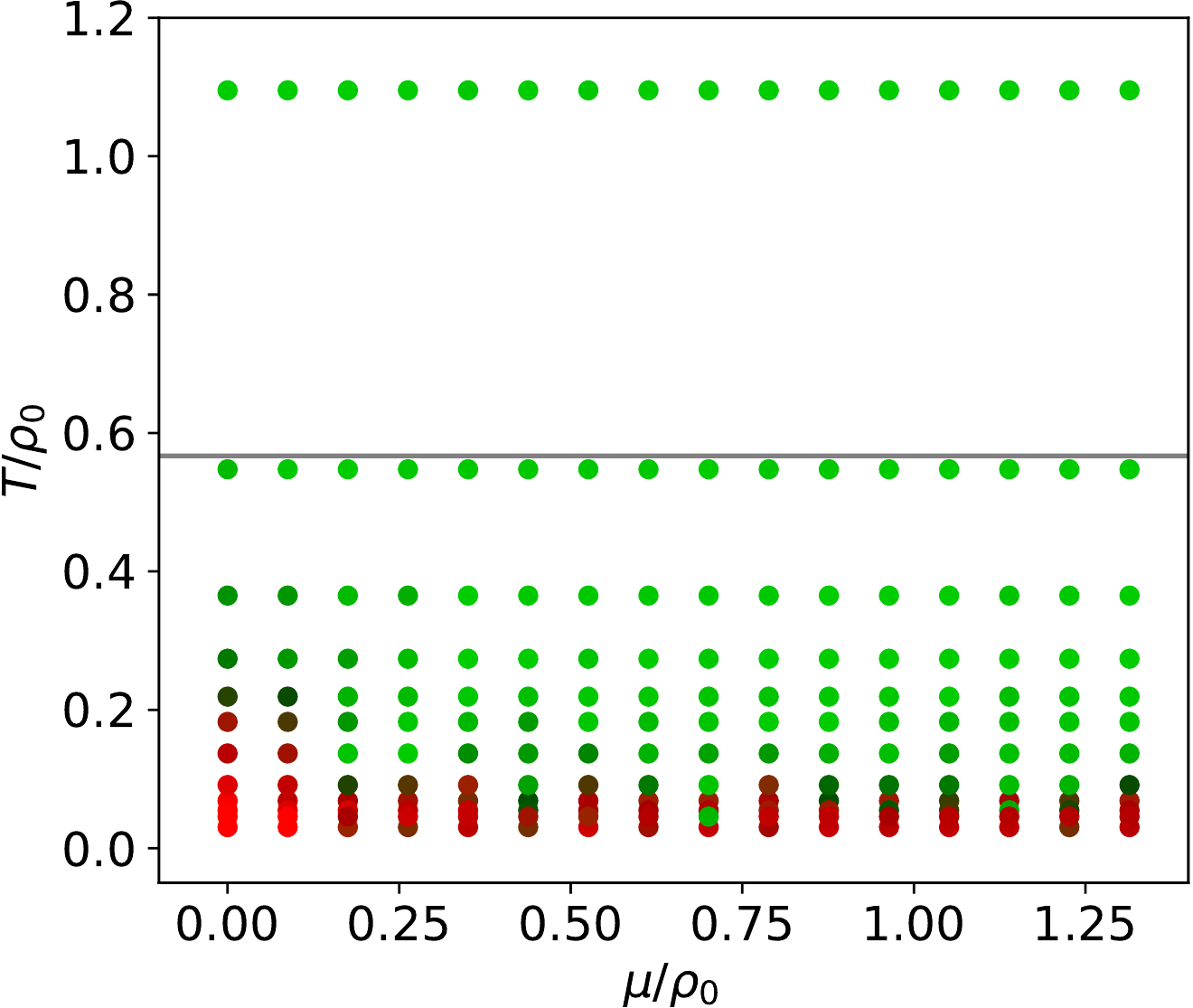}
\end{subfigure}
\hspace{.11cm}
\begin{subfigure}{.333\linewidth}
	\includegraphics[scale=0.33]{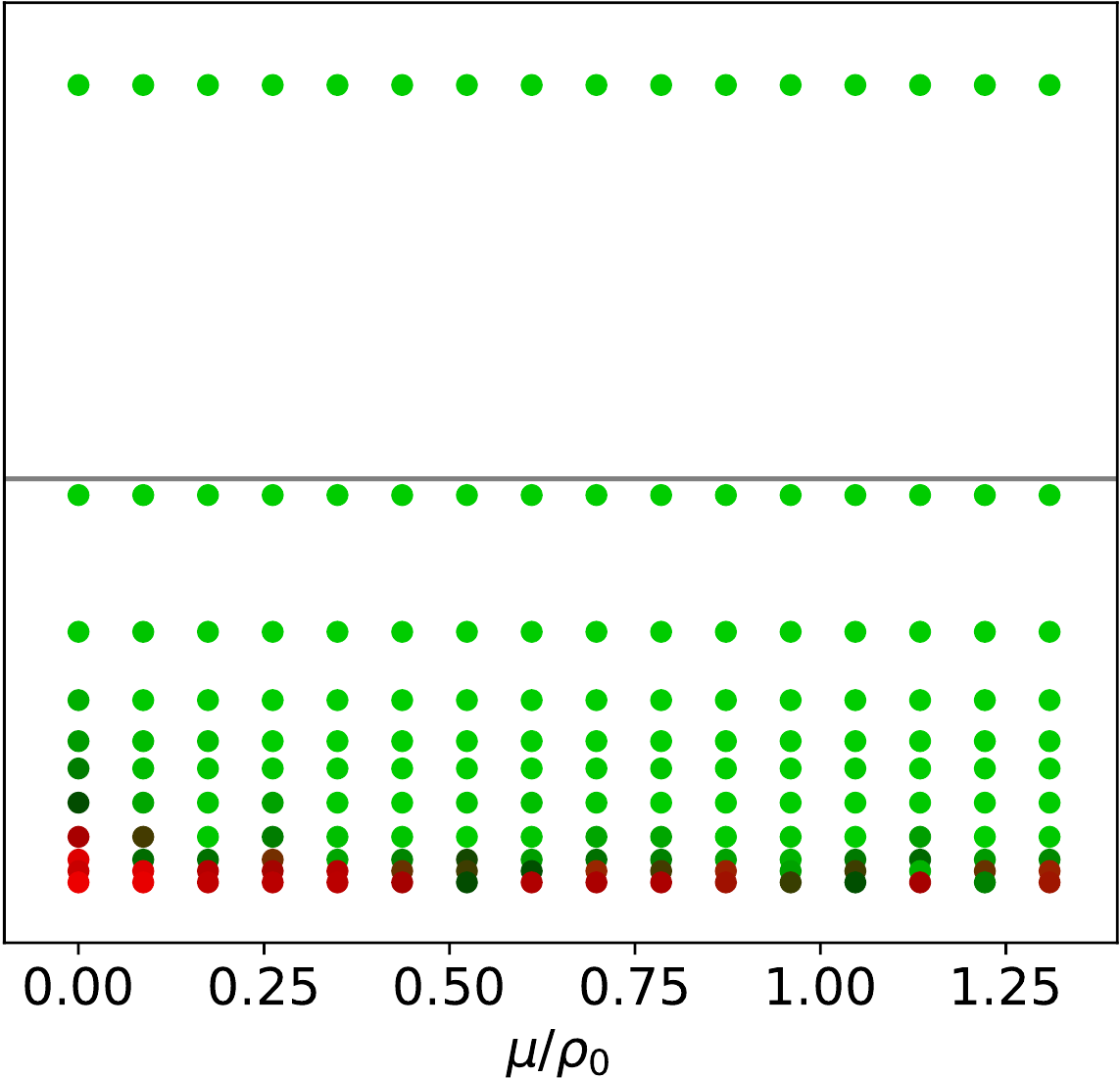}
\end{subfigure}
\hspace{-.4cm}
\begin{subfigure}{.333\linewidth}
	\includegraphics[scale=0.33]{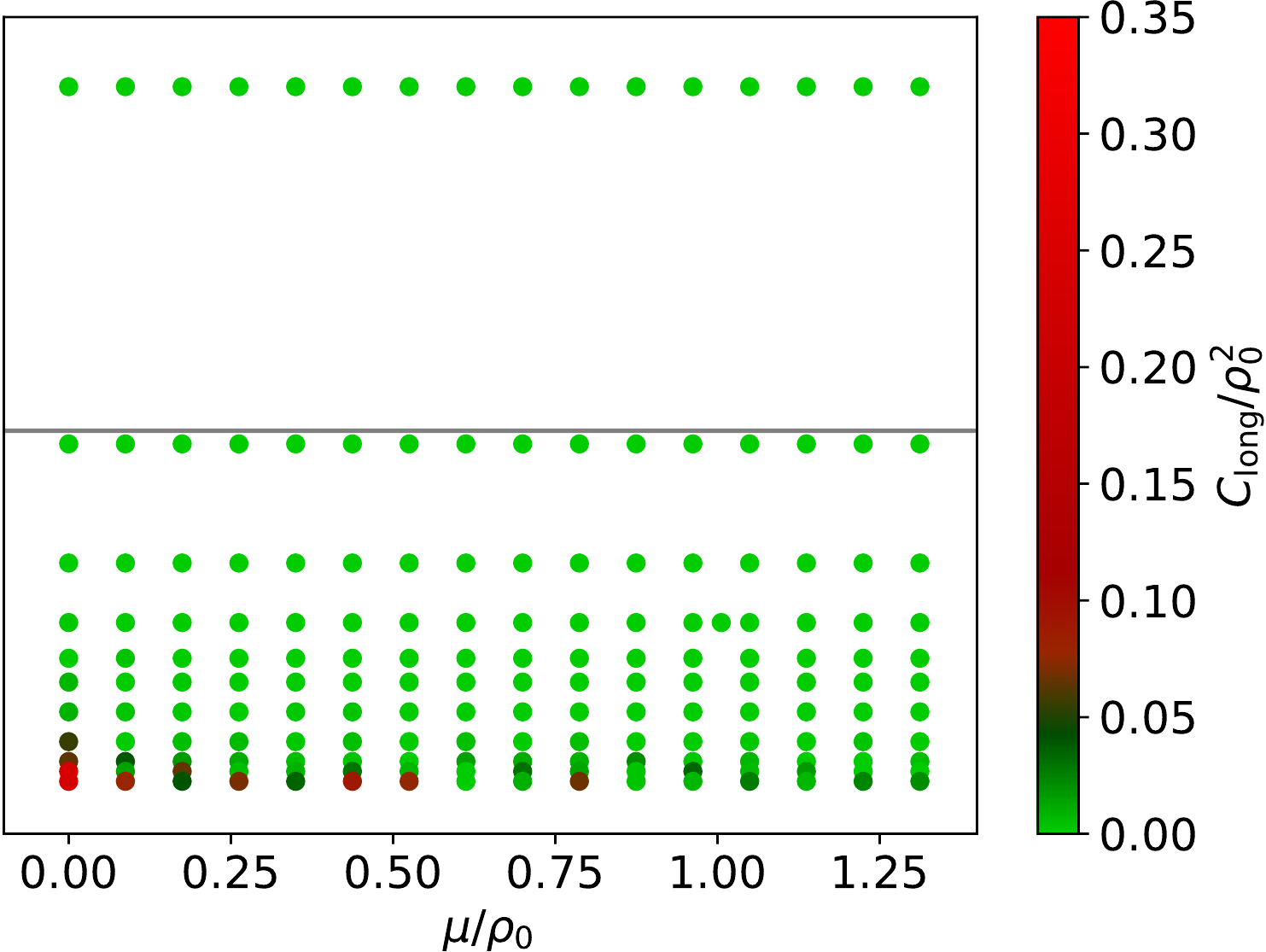}
\end{subfigure}
\caption{From left to right: infinite-volume extrapolation of the $(\mu,T)$ phase 
	diagram at fixed lattice spacing; top row: using $C_{\mathrm{short}}$ 
	(Eq.~\eqref{eq:cshort}, taken from \cite{LMW21}) ; bottom row: using 
	$C_{\mathrm{long}}$ (Eq.~\eqref{eq:clong}).}
\label{fig:pd_infinite_volume}
\end{figure}

We study phase diagrams in the $(\mu,T)$ plane using
\eqref{eq:cshort} and \eqref{eq:clong}. In order to quantify finite-volume effects 
we performed an infinite-volume extrapolation at fixed lattice spacing and show the 
results in Fig.~\ref{fig:pd_infinite_volume}. 

First of all, there is a region where  $\Cs>0$ and thus homogeneous 
configurations dominate, i.e. the red region in Fig.~\ref{fig:pd_infinite_volume}. 
Its extent in the $\mu$-direction shrinks in the infinite-volume limit, which can 
be understood by recalling that the chiral spiral's wavelength is inversely 
proportional to $\mu$. This means that for small but non-vanishing chemical potential 
smaller lattices are simply not large enough to fit in a full wavelength, thus 
leading to a suppression of inhomogeneous configurations. The large extent of the 
red region in the $T$-direction on the smallest lattice can also be seen to be a 
finite-size effect.

We furthermore observe that the boundary between the "short-ranged" 
($\Cs\approx0$) and "inhomogeneous" ($\Cs<0$) regions remains essentially unchanged 
in the infinite-volume extrapolation. This is actually expected because the
appearance of $\Cs<0$ only depends on the two scales introduced by the temperature
and the chemical potential respectively, neither of which changes in the infinite-volume limit. 
%(or continuum limit later on). 
More precisely, negative values of $\Cs$ occur whenever the preferred
wavelength of the oscillatory (quasi-)condensates is smaller than the thermally
induced finite correlation length. From their analytically known values in an
expansion in $1/\Nf$ (see \cite{LMW21}), it is easy to anticipate the shape of this
boundary -- at least on a qualitative level.
In contrast, $\Cl$ becomes smaller and smaller in the infinite-volume limit since it
probes the largest scales on which we do not expect correlations.
This is indeed strong evidence against spontaneous symmetry breaking
and fundamentally differs from what was observed in \cite{LPW20}.

\begin{figure}
\begin{subfigure}{.333\linewidth}
	\includegraphics[scale=0.33]{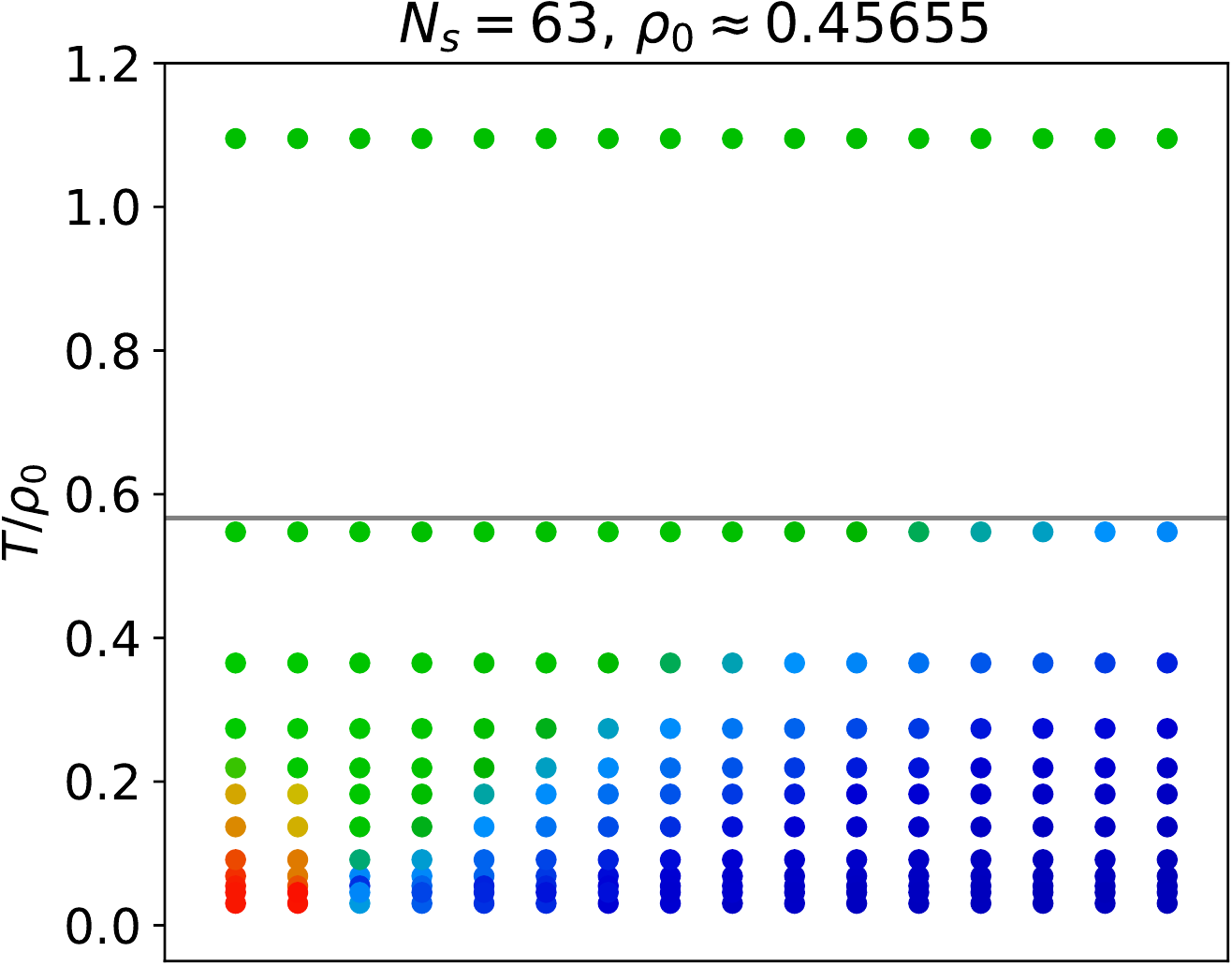}
\end{subfigure}
\hspace{-2.3cm}
\begin{subfigure}{.333\linewidth}
	\includegraphics[scale=0.33]{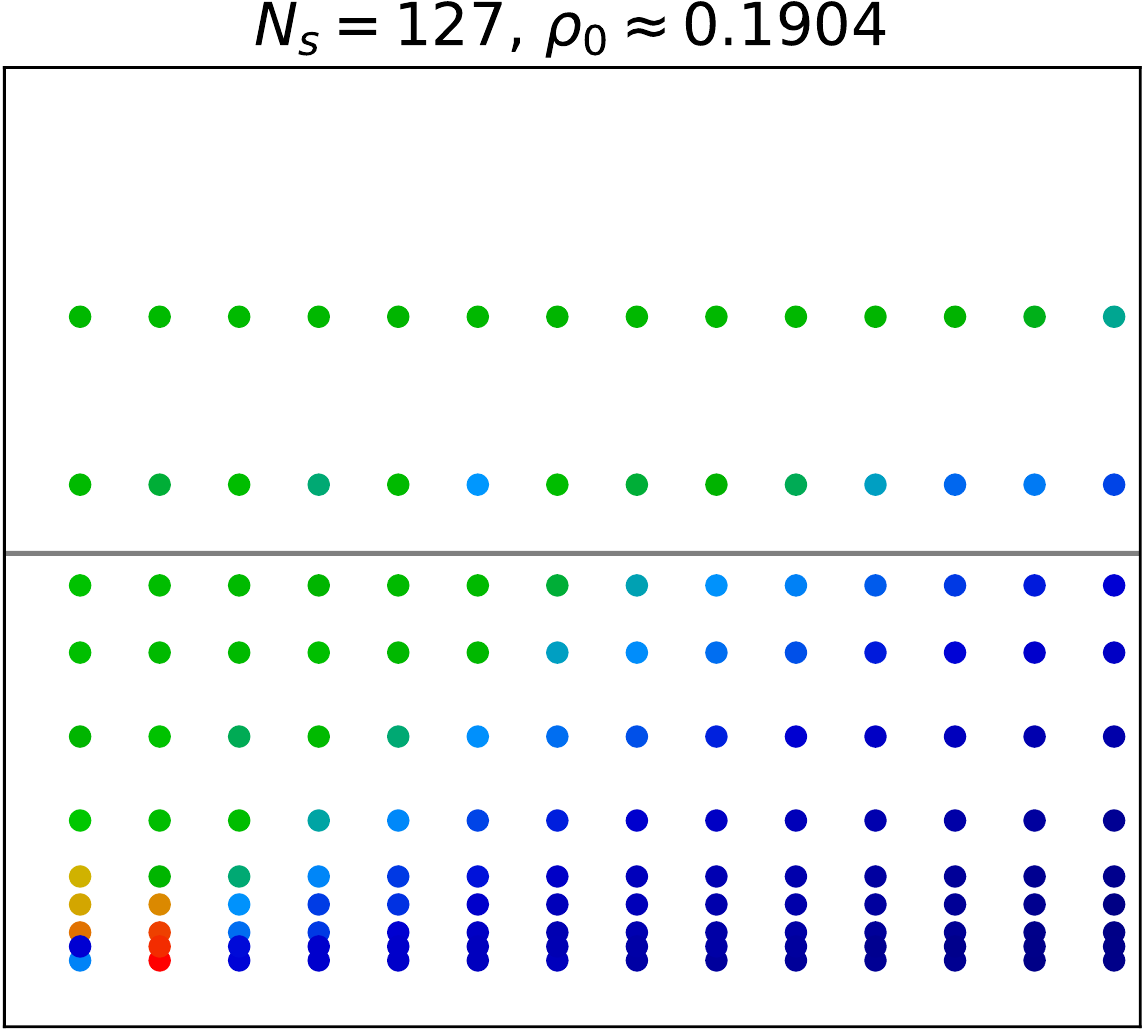}
\end{subfigure}
\hspace{-2.8cm}
\begin{subfigure}{.333\linewidth}
	\includegraphics[scale=0.33]{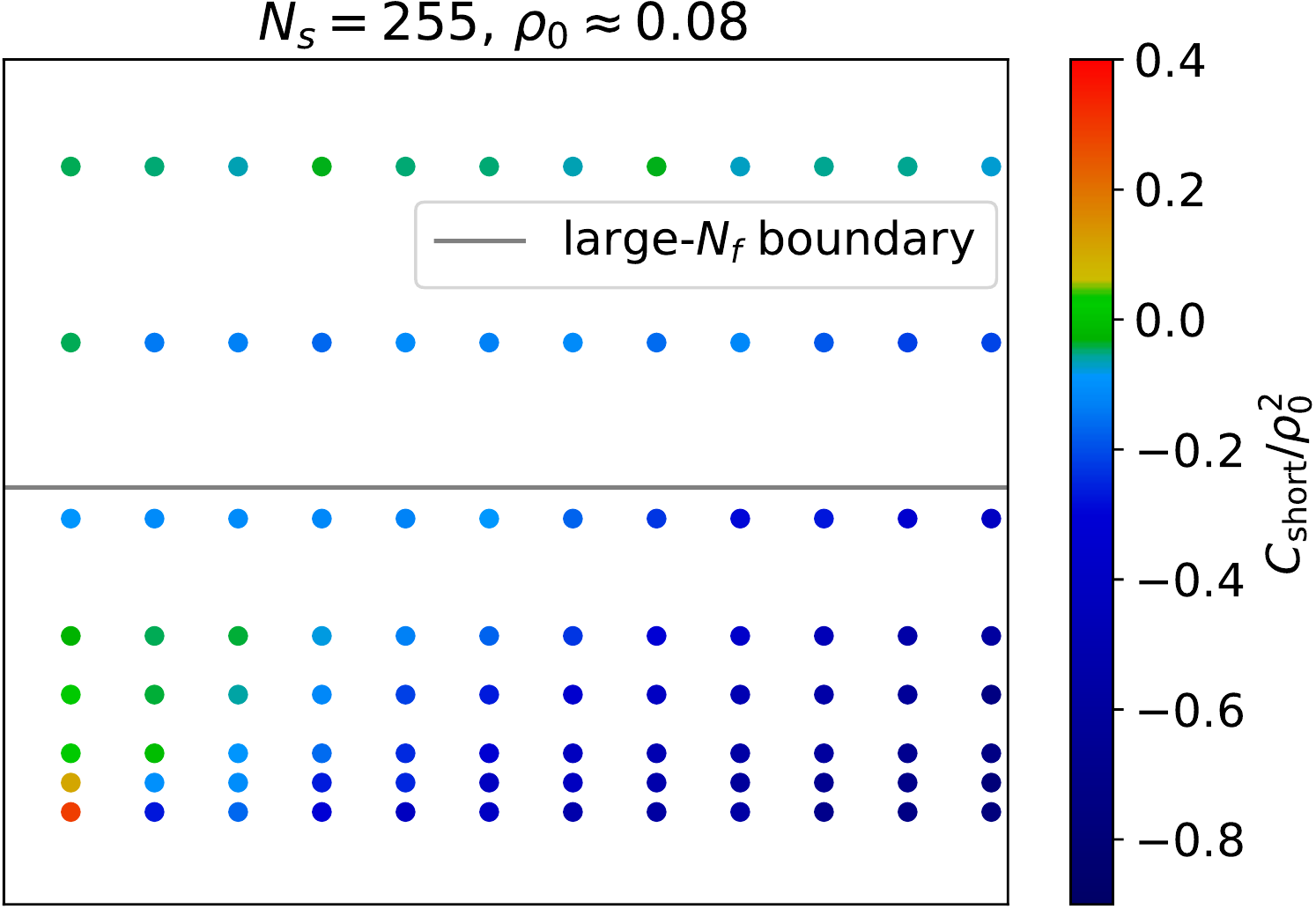}
\end{subfigure}
\begin{subfigure}{.333\linewidth}
	\includegraphics[scale=0.33]{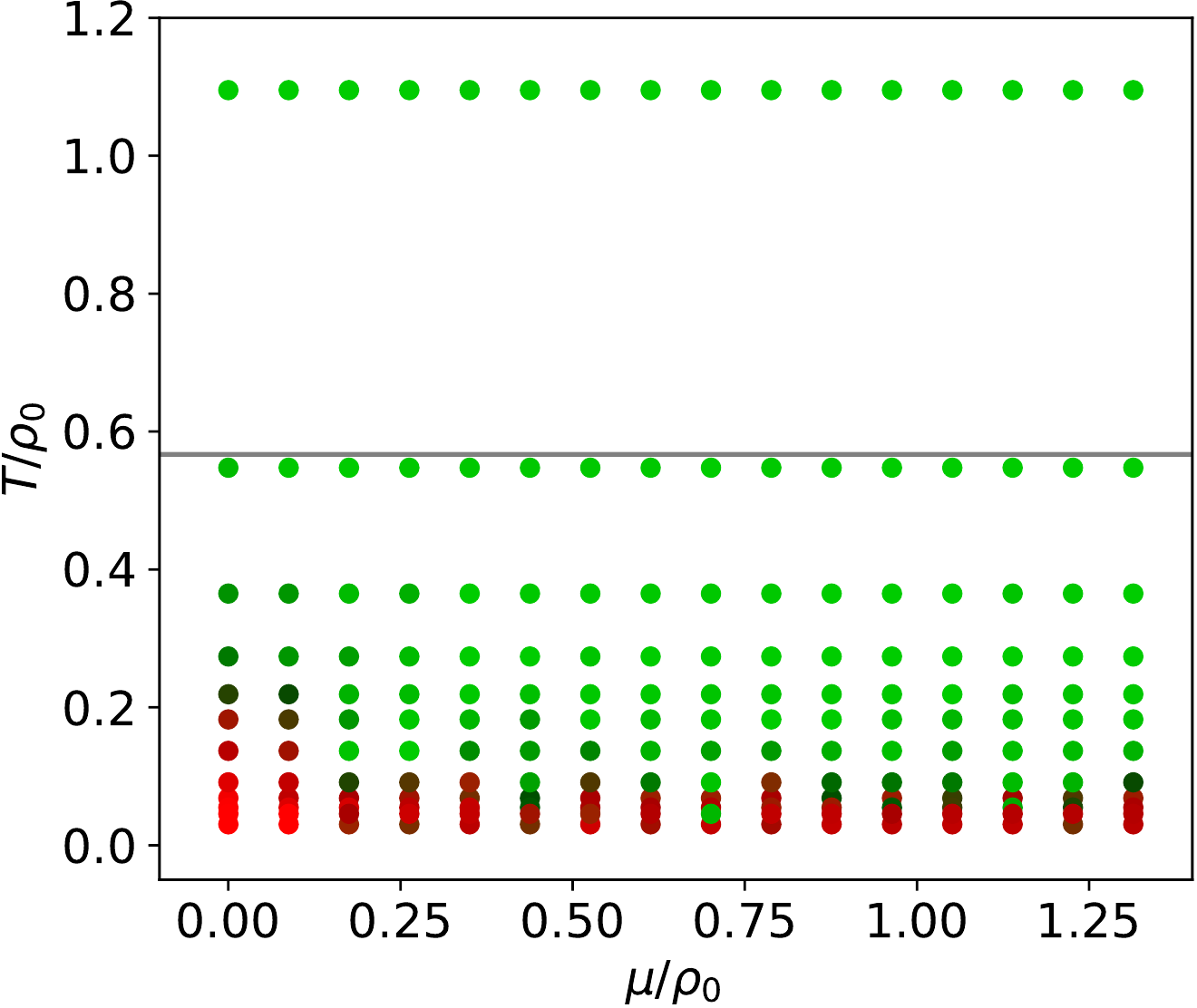}
\end{subfigure}
\hspace{.11cm}
\begin{subfigure}{.333\linewidth}
	\includegraphics[scale=0.33]{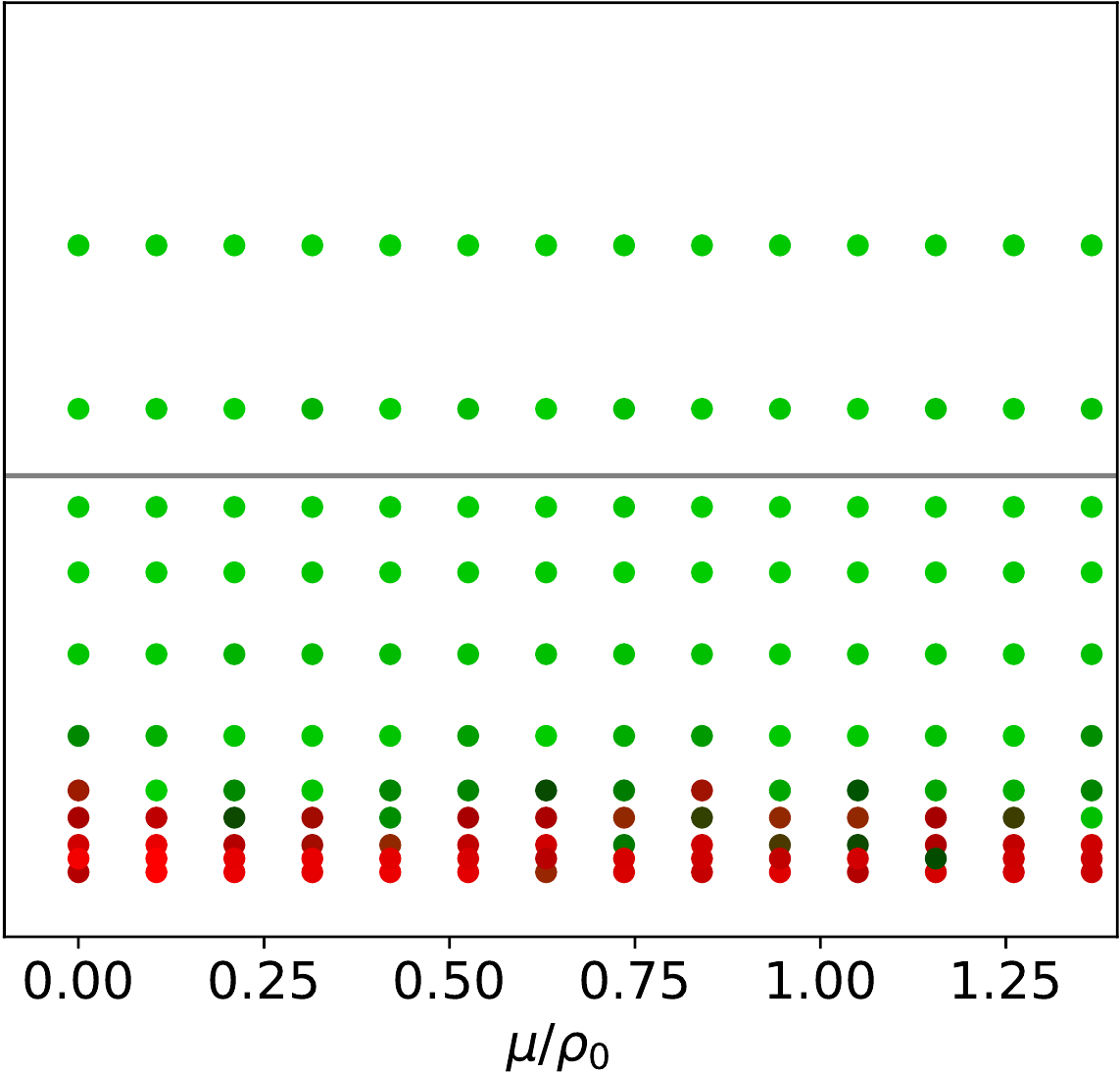}
\end{subfigure}
\hspace{-.4cm}
\begin{subfigure}{.333\linewidth}
	\includegraphics[scale=0.33]{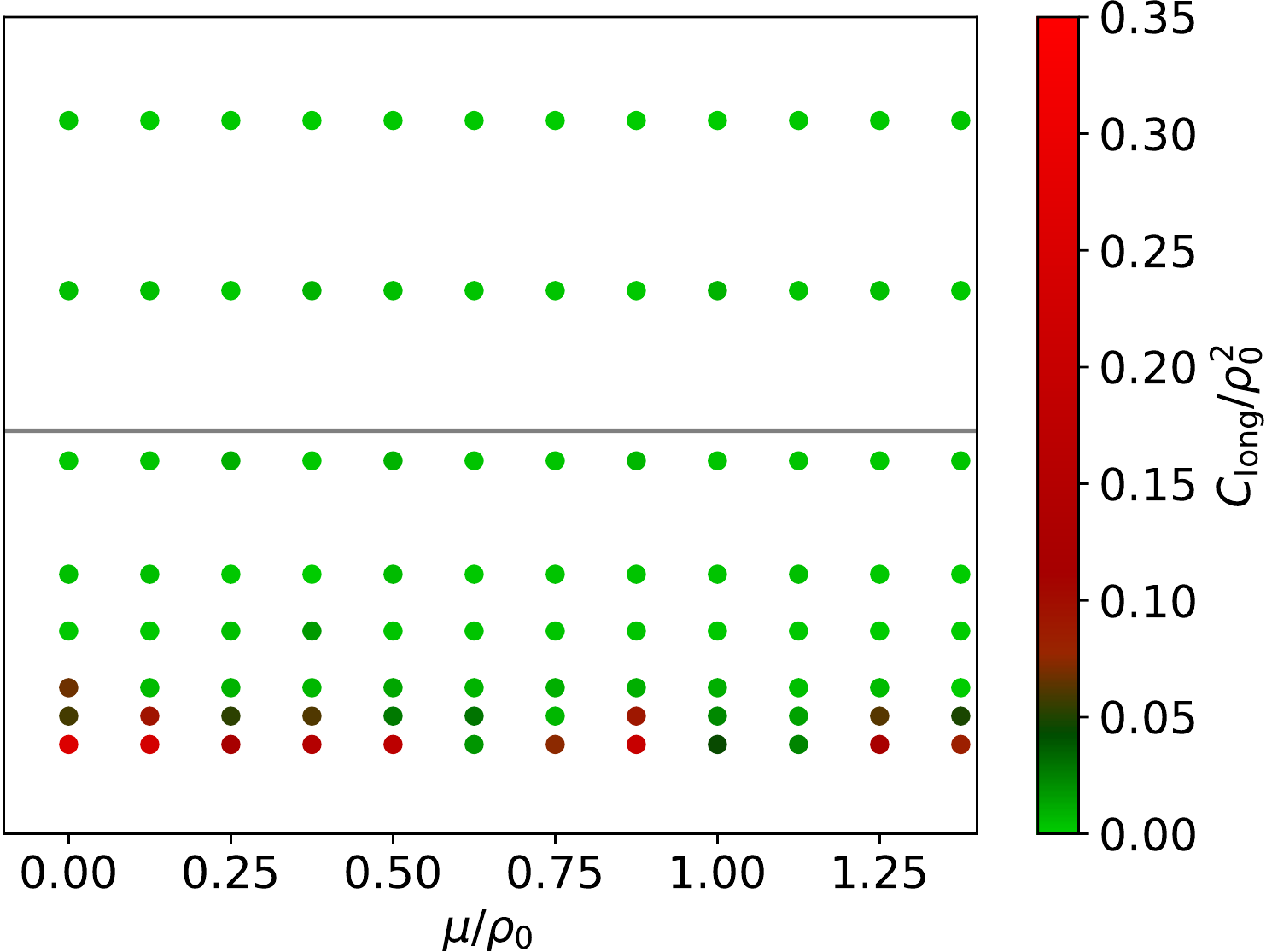}
\end{subfigure}
\caption{From left to right: continuum extrapolation of the $(\mu,T)$ phase 
	diagram at roughly fixed physical volume; top row: using $C_{\mathrm{short}}$ 
	(Eq.~\eqref{eq:cshort}, taken from \cite{LMW21}); bottom row: using 
	$C_{\mathrm{long}}$ (Eq.~\eqref{eq:clong}).}
\label{fig:pd_continuum}
\end{figure}

Finally, we study discretization effects via a continuum extrapolation at roughly 
fixed physical volume in Fig.~\ref{fig:pd_continuum}. We find remnant structural order
for three strongly decreasing lattice spacings. However, due to subtleties in the
scale-setting procedure (see \cite{LMW21}) we refrain from drawing 
definite conclusions at this point and instead hope to revisit the continuum limit
in the future. In related models in higher dimensions the continuum limit turned out
to be crucial for removing misleading lattice artifacts \cite{BKW21}. We do emphasize,
however, that our results are consistent with the
persistence of remnant inhomogeneities as $a$ appproaches zero -- in accordance with
the observation that in lower dimensions the UV limit usually tends to be less problematic.

\clearpage

\newpage
\acknowledgments

We are indebted to Laurin Pannullo, Marc Wagner, Marc Winstel and 
Andreas Wipf for numerous enlightening discussions and previous
collaborations on inhomogeneous condensates. We furthermore thank 
Björn Wellegehausen for providing the simulation code base used in 
this work. This work has been funded by the Deutsche 
Forschungsgemeinschaft (DFG) under Grant No. 406116891 within the 
Research Training Group RTG 2522/1. The numerical simulations were 
performed on resources of the Friedrich Schiller University of Jena 
supported in part by the DFG grants INST 275/334-1 FUGG and INST 
275/363-1 FUGG. 

\bibliographystyle{JHEP}
\bibliography{bibliography}

\end{document}